\RequirePackage{ifpdf}
\ifpdf 
\documentclass[pdftex]{sigma}
\else
\documentclass{sigma}
\fi

\begin{document}

\allowdisplaybreaks

\renewcommand{\thefootnote}{$\star$}

\renewcommand{\PaperNumber}{041}

\FirstPageHeading

\ShortArticleName{Solutions of Zakharov--Yajima--Oikawa System}

\ArticleName{Periodic and Solitary Wave Solutions \\ of Two Component
Zakharov--Yajima--Oikawa System,\\ Using Madelung's Approach\footnote{This
paper is a contribution to the Proceedings of the Conference ``Symmetries and Integrability of Dif\/ference Equations (SIDE-9)'' (June 14--18, 2010, Varna, Bulgaria). The full collection is available at \href{http://www.emis.de/journals/SIGMA/SIDE-9.html}{http://www.emis.de/journals/SIGMA/SIDE-9.html}}}

\Author{Anca VISINESCU~$^\dag$, Dan GRECU~$^\dag$, Renato FEDELE~$^\ddag$ and Sergio DE NICOLA~$^\S$}
\AuthorNameForHeading{A.~Visinescu, D.~Grecu,  R.~Fedele and S.~De Nicola}

\Address{$^\dag$~Department of Theoretical Physics, National Institute for Physics and Nuclear Engineering, \\
\hphantom{$^\dag$}~Bucharest, Romania}
\EmailD{\href{mailto:avisin@theory.nipne.ro}{avisin@theory.nipne.ro}, \href{mailto:dgrecu@theory.nipne.ro}{dgrecu@theory.nipne.ro}}

\Address{$^\ddag$~Dipartimento di Scienze Fisiche, Universita Federico II and INFN Sezione di
Napoli,\\
\hphantom{$^\ddag$}~Napoli, Italy}
\EmailD{\href{mailto:renato.fedele@na.infn.it}{renato.fedele@na.infn.it}}

 \Address{$^\S$~Istituto Nazionale di Ottica del Consiglio Nazionale delle Ricerche,  Pozuolli, (Na), Italy}
 \EmailD{\href{mailto:sergio.denicola@ino.it}{sergio.denicola@ino.it}}

\ArticleDates{Received February 10, 2011, in f\/inal form April 19, 2011;  Published online April 23, 2011}

\Abstract{Using the multiple scales method, the interaction
between two bright and one dark solitons is studied. Provided that
a long wave-short wave resonance condition is sa\-tisf\/ied, the
two-component Zakharov--Yajima--Oikawa (ZYO) completely
integrable system is obtained. By using a Madelung f\/luid
description, the one-soliton solutions of the corresponding ZYO
system are determined. Furthermore, a discussion on the
interaction between one bright and two dark solitons is presented.
In particular, this problem is reduced to solve a one-component
ZYO system in the resonance conditions.}


\Classification{35Q55; 37K10; 45G15}

\section{Introduction}

In many physical applications more than one single wave is
propagating in a nonlinear medium, and the interaction of several
waves has to be taken into account. In particular, to describe the
propagation of three nonlinear pulses in some dispersive material
one has to solve simultaneously a set of coupled nonlinear
Schr\"{o}dinger (NLS) equations. For instance, it occurs in: (i)~the propagation of solitonlike pulses in birefringent optical
f\/ibers \cite{KA,T,M,CEEK}; (ii)~the nonlinear wave dynamics of
Bose--Einstein condensates~\cite{CFK,PBKK}; (iii)~the
soliton propagation through optical f\/iber array \cite{KA,A,AA,HK}; (iv)~the nonlinear dynamics of gravity waves
in crossing sea states~\cite{OOS}.

In this paper, a theoretical investigation on the interaction
between two bright solitons (short waves) and a dark soliton (long
wave) is carried out on the basis of previous investigations that
have reduced, under suitable conditions, the study of bright-dark
soliton interaction to the study of the existence of a ``long
wave-short wave resonance'' (LW-SW resonance)~\cite{Kiv}. This
resonance phenomena has quite a large universality. For instance,
in plasma physics it describes Langmuir solitons moving near the
speed of sound \cite{Zak,YO,LS}, in hydrodynamics it appears
in the study of internal gravity waves~\cite{G} and a general
study of LW-SW resonance \cite{B,DR}, in
quasi-one-dimensional molecular crystals it describes the
resonance between the exciton and phonon f\/ields in Davydov's model
\cite{D,DEGM,BLPS,VG}. Recent extensions of the LW-SW resonance to
two dimensions and more components have been discussed and solved
by several authors \cite{RKLT,Y,OMO}.

In the resonance condition, we try to reduce our problem to the
study of the completely integrable system of two-component
Zakharov--Yajima--Oikawa \cite{Zak,YO} (see equations~(9) and~(10) of~\cite{Kiv}). Besides its relevance in nonlinear optics,
the same system describes the Davydov model with two excitonic
components \cite{GGV}. It is interesting to mention that a vector-like generalization of Zakharov--Benney's equations for long-short wave interaction were obtained 25 years ago in the study of magnon-phonon interaction in a many sublattice XY chain~\cite{MPK}, which in the case of one-sublattice XY chain reduces to the Yajima--Oikawa system.  In the next section, the basic equations
describing the three wave interaction will be presented and, using
a multiple scales analysis, the Zakharov--Yajima--Oikawa system is
obtained. In Section~\ref{section3}, the Madelung f\/luid description is used to
discuss analytically the above system. In particular, several
solitary solutions in the form of traveling waves are presented.
In Section~\ref{section4}, the interaction between one-bright and two-dark
solitons is discussed, and a simple one-component
Zakharov--Yajima--Oikawa system is obtained in resonance
condition. In particular, one soliton solution is presented.
Finally, remarks and conclusion are presented in Section~\ref{section5}.

\section{Basic equations and multiple scales analysis}\label{section2}

We consider three nonlinear dispersive waves propagating in an
optical f\/iber. We assume that these waves are associated with
weakly nonlinear dispersion relations, here denoted by $
\omega_{j}=\omega_{j}( k_{j}: \vert A_{1} \vert^{2}, \vert
A_{2}\vert^{2}, \vert A_{3}\vert^{2})$, $j=1,2,3 $, respectively.
Each $A_j$ stands for a complex amplitude that, due to the weakly
nonlinear dynamics of the medium, is af\/fected by a modulation in
both space and time. As it is well known \cite{Whitham}, due to
this dynamics, each mode of the system can be represented by a
wave packet where a carrier wave is amplitude modulated. In order
to obtain the evolution equation for each mode, we apply the
well-known method of Karpman and Kruskal~\cite{KK} (see also
\cite{Whitham,Karpman}). To this end, let us denote by $
e^{i(k_{0}x-\omega_{0}t)}$ the basic carrier wave. Then a
Taylor expansion around $(k_{0},\omega_{0})$ and $\vert A_{i}
\vert =0 $ of each $\omega_{i}$ will give
\begin{gather*}
\omega_{i}-\omega_{0}=\left(\frac{\partial \omega_{i}}{\partial
k_{i}}\right)_{0}(k_{i}-k_{0})+\frac{1}{2}\left(\frac{\partial^{2}\omega_{i}}{\partial
k^{2}_{i}}\right)_{0}(k_{i}-k_{0})^{2}+
\left(\frac{\partial\omega_{i}}{\partial \vert
A_{1}\vert^{2}}\right)_{0}\vert A_{1}\vert^{2}\\
\phantom{\omega_{i}-\omega_{0}=}{} +
\left(\frac{\partial\omega_{i}}{\partial \vert
A_{2}\vert^{2}}\right)_{0}\vert A_{2}\vert^{2}+
\left(\frac{\partial\omega_{i}}{\partial \vert
A_{3}\vert^{2}}\right)_{0}\vert A_{3}\vert^{2}+\cdots.
\end{gather*}
Replacing  $\omega_{i}-\omega_{0} \simeq
-i\frac{\partial}{\partial t}$, $k_{i}-k_{0}\simeq
i\frac{\partial}{\partial x}$, after a translation of coordinate
$\big(x\rightarrow x-\big(\frac{\partial \omega_{3}}{\partial
k_{3}}\big)_{0}t\big)$, the following nonlinear system of three
interacting waves is obtained
\begin{gather}
i\frac{\partial A_{1}}{\partial t}+iV_{1}\frac{\partial A_{1}}{\partial x}+\frac{\alpha_{1}}{2}\frac{\partial^{2}A_{1}}{\partial x^{2}}+\alpha_{2}\vert A_{1}\vert^{2}A_{1}+\alpha_{3}\vert A_{2}\vert^{2}A_{1}+\alpha_{4}\vert A_{3}\vert^{2}A_{1}=0,\nonumber \\
i\frac{\partial A_{2}}{\partial t}+iV_{2}\frac{\partial A_{2}}{\partial x}+\frac{\beta_{1}}{2}\frac{\partial^{2}A_{2}}{\partial x^{2}}+\beta_{2}\vert A_{1}\vert^{2}A_{2}+\beta_{3}\vert A_{2}\vert^{2}A_{2}+\beta_{4}\vert A_{3}\vert^{2}A_{2}=0,\nonumber\\
i\frac{\partial A_{3}}{\partial
t}+\frac{\gamma_{1}}{2}\frac{\partial^{2}A_{3}}{\partial
x^{2}}+\gamma_{2}\vert A_{1}\vert^{2}A_{3}+\gamma_{3}\vert
A_{2}\vert^{2}A_{3}+\gamma_{4}\vert
A_{3}\vert^{2}A_{3}=0.\label{2}
\end{gather}
Here we denoted $ V_{i}=\left(\frac{\partial \omega_{i}}{\partial
k_{i}}\right)_{0}-\left(\frac{\partial \omega_{3}}{\partial
k_{3}}\right)_{0}$, $i=1,2 $ and the constants  $
\alpha_{1}$, $\beta_{1}$, $\gamma_{1} $  are related to derivatives of
$\omega_{i}$  with respect to  $k_{i}$ (ex.
$\alpha_{1}=-\left(\frac{\partial^{2}\omega_{1}}{\partial
k_{1}^{2}}\right)_{0},\cdot$)  while $ \alpha_{2},\dots,\gamma_{4} $ to
the derivatives with respect to $ \vert A_{i}\vert^{2}$ (ex. $
\alpha_{2}=\left(\frac{\partial \omega_{1}}{\partial \vert
A_{1}\vert^{2}}\right)\cdots$).

Further on we shall consider channel~3 with normal dispersion and~1 and 2 with the anomalous one~\cite{KA}. Then following Kivshar~\cite{Kiv} it is convenient to introduce new f\/ield variables
\begin{gather*}
A_{1}=\Psi_{1}e^{i\delta_{1}t}, \qquad A_{2}=\Psi_{2}e^{i\delta_{2}t},\qquad A_{3}=(u_{0}+a(x,t))e^{i(\Gamma
t+\phi(x,t))},
\\
\delta_{i}=\left(\frac{\partial \omega_{i}}{\partial \vert
A_{3}\vert^{2}}\right)_{0}u^{2}_{0},
\qquad \Gamma=\left(\frac{\partial\omega_{3}}{\partial\vert
A_{3}\vert^{2}}\right)_{0}u^{2}_{0}
\end{gather*}
($u_{0}$, $a(x,t)$, $\phi(x,t)$ being real quantities). Note that for
solitary wave solutions the additional conditions of vanishing of
$a$ at the inf\/inity has to be imposed. Then the equations of
$A_{1}$ and $A_{2}$ become, respectively
\begin{gather}
i\frac{\partial \Psi_{1}}{\partial t}+iV_{1}\frac{\partial
\Psi_{1}}{\partial x}+\frac{\alpha_{1}}{2}\frac{\partial^{2}\Psi_{1}}{\partial x^{2}}+
\left(\alpha_{2}\vert \Psi_{1}\vert^{2}+\alpha_{3}\vert \Psi_{2}\vert^{2}\right)\Psi_{1}+
2\alpha_{4}u_{0}a\Psi_{1}+\alpha_{4}a^{2}\Psi_{1}=0, \nonumber \\
i\frac{\partial \Psi_{2}}{\partial t}+iV_{2}\frac{\partial
\Psi_{2}}{\partial
x}+\frac{\beta_{1}}{2}\frac{\partial^{2}\Psi_{2}}{\partial
x^{2}}+\left(\beta_{2}\vert \Psi_{1}\vert^{2}+\beta_{3}\vert
\Psi_{2}\vert^{2}\right)\Psi_{2}+2\beta_{4}u_{0}a\Psi_{2}+\beta_{4}a^{2}\Psi_{2}=0.\label{3}
\end{gather}
As concerns the $A_{3}$-equation, separating the real and the
imaginary part, the following system of coupled equations is
obtained
\begin{gather*}
\frac{\partial a}{\partial
t}+\frac{\gamma_{1}}{2}u_{0}\frac{\partial^{2}\phi}{\partial
x^{2}}+\mbox{(nonlinear~terms)}=0,
\\
-\frac{\partial\phi}{\partial
t}+2\gamma_{4}u_{0}a+\frac{\gamma_{1}}{2u_{0}}\frac{\partial^{2}a}{\partial
x^{2}}+\left(\gamma_{2}\vert\Psi_{1}\vert^{2}+\gamma_{3}\vert\Psi_{2}\vert^{2}\right)+\mbox{(nonlinear~terms)}=0.
\end{gather*}
In both these equations the parenthesis $(\cdots)$ contains all the
other nonlinear terms which will be irrelevant in a multiple
scales analysis. From these last two equations the following
equation satisf\/ied by $a(x,t)$ is easily obtained
\begin{gather}\label{4}
\frac{\partial^{2}a}{\partial
t^{2}}+\gamma_{1}\gamma_{4}u^{2}_{0}\frac{\partial^{2}a}{\partial
x^{2}}+\frac{\gamma_{1}^{2}}{4}\frac{\partial^{4}a}{\partial
x^{4}}+u_{0}\frac{\gamma_{1}}{2}\frac{\partial^{2}}{\partial
x^{2}}\left(\gamma_{2}\vert \Psi_{1}\vert^{2}+\gamma_{3}\vert
\Psi_{2}\vert^{2}\right)
\\
\qquad {}+\mbox{(higher~order~nonlinear  terms   in  $(a,\phi)$ and~their~derivatives)}=0.
\nonumber
\end{gather}

The linear part of the $a$ equation corresponds to an acoustic
f\/ield with dispersion relation ($\gamma_{1}<0$, $\gamma_{4}>0$)
\[
\omega =ck\sqrt{1+\frac{\gamma_{1}^{2}}{4c^{2}}k^{2}}\simeq
ck\left(1+\frac{\gamma_{1}^{2}}{8c^{2}}k^{2}\right)
\]
and phase velocity $c=\omega /k$, where
$c^{2}=\vert\gamma_{1}\vert \gamma_{4}u^{2}_{0}.$

We shall perform a multiple scales analysis of the system
\eqref{3}~+~\eqref{4} \cite{Kiv}. We introduce new scaled variables
\[
t\Rightarrow\epsilon t, \qquad x\Rightarrow\sqrt{\epsilon}(x-ct)
\]
and new functions
\[
a\Rightarrow\epsilon a,~~\phi\Rightarrow\epsilon\phi,
\qquad \Psi_{1}\Rightarrow\epsilon^{\frac{3}{4}}\Psi_{1},
\qquad \Psi_{2}\Rightarrow\epsilon^{\frac{3}{4}}\Psi_{2}.
\]
Then in order $\frac{5}{2}$ in $\epsilon$ from $a$  equation we
obtain
\begin{gather}\label{5}
-2c\frac{\partial a}{\partial
t}+u_{0}\frac{\gamma_{1}}{2}\frac{\partial}{\partial
x}\left(\gamma_{2}\vert\Psi_{1}\vert^{2}+
\gamma_{3}\vert\Psi_{2}\vert^{2}\right)=0.
\end{gather}
All the other terms in $a$ equation contribute to higher order in
$\epsilon$. In the order  $\frac{5}{4}$ from $\Psi_{i}$ equations
we obtain $V_{1}=V_{2}=c. $ This is the well known long wave-short
wave (LW-SW) resonance condition: ``{\it  the group velocity $V$ of
the SW is equal to the phase velocity of the LW}''~\cite{B}. In the
next order $\left(\frac{7}{4}\right)$ in $\epsilon $ from the
$\Psi$ equations we get
\begin{gather}
i\frac{\partial \Psi_{1}}{\partial
t}+\frac{\alpha_{1}}{2}\frac{\partial^{2}\Psi_{1}}{\partial
x^{2}}+2\alpha_{4}u_{0}a\Psi_{1}=0,\nonumber\\
\label{6}
i\frac{\partial \Psi_{2}}{\partial
t}+\frac{\beta_{1}}{2}\frac{\partial^{2}\Psi_{2}}{\partial
x^{2}}+2\beta_{4}u_{0}a\Psi_{2}=0.
\end{gather}
The equations \eqref{5}~+~\eqref{6} represent an 1-dimensional
2-components Zakharov \cite{Zak}, Yajima--Oikawa~\cite{YO} system.
As mentioned in the Introduction the same system in the same LW-SW
resonance condition was obtained in a Davydov model with two
excitonic modes coupled with a phonon f\/ield~\cite{GGV}. The same
line of reasoning was used in~\cite{OMO} for three interacting
waves in 2-dimensions.

\section{Madelung f\/luid description}\label{section3}

The special case $(\alpha_{i}=\beta_{i}$, $\gamma_{2}=\gamma_{3})$
is  completely integrable \cite{OMO}  and will be considered in
the following. In this case, simplifying the notations, the system
\eqref{5}~+~\eqref{6}  is written in the following form
($\gamma>0$, $\beta>0$)
\begin{gather}
\frac{\partial a}{\partial t}-\gamma\frac{\partial}{\partial x}\left(\vert\Psi_{1}\vert^{2}+\vert\Psi_{2}\vert^{2}\right)=0, \nonumber \\
i\frac{\partial\Psi_{i}}{\partial
t}+\frac{1}{2}\frac{\partial^{2}\Psi_{i}}{\partial
x^{2}}-\beta\Psi_{i}a=0, \qquad i=1,2.\nonumber
\end{gather}
The $\Psi_{i}$ equations will be transformed using a Madelung
f\/luid description \cite{Mad,Fed}. We write
\[
\Psi_{i}=\sqrt{\rho_{i}} e^{i \theta_{i}},
\]
where $\rho_{i}$, $\theta_{i}$ are real functions of $(x,t)$ and
moreover $\rho_{i}$ are positive quantities. Introducing this
expression into $a$-equation this becomes
\begin{gather}\label{8}
\frac{\partial a}{\partial t}-\gamma\frac{\partial}{\partial
x}(\rho_{1}+\rho_{2})=0,
\end{gather}
while from the $\Psi_{i}$ equations, after the separation of real
and imaginary parts, we obtain
\begin{gather*}
 \frac{\partial \rho_{i}}{\partial t}+\frac{\partial}{\partial x}(v_{i}\rho_{i})=0,
\end{gather*}
which is a continuity equation for the f\/luid densities
$\rho_{i}=\vert\Psi_{i}\vert^{2}$ with $ v_{i}(x,t)=\frac{\partial
\theta_{i}(x,t)}{\partial x}$
the f\/luid velocities components and
\begin{gather}\label{10}
-\frac{\partial\theta_{i}}{\partial
t}+\frac{1}{2}\frac{1}{\sqrt{\rho_{i}}}\frac{\partial^{2}\sqrt{\rho_{i}}}{\partial
x^{2}}-\frac{1}{2}\left(\frac{\partial\theta_{i}}{\partial
x}\right)^{2}-\beta a=0.
\end{gather}
Dif\/ferentiating this last expression with respect to $x$ the
following equations of motion for the f\/luid velocities $v_{i}$ are
obtained
\begin{gather}\label{11}
\left(\frac{\partial}{\partial t}+v_{i}\frac{\partial}{\partial
x}\right)v_{i}=\frac{1}{2}\frac{\partial}{\partial
x}\left(\frac{1}{\sqrt{\rho_{i}}}\frac{\partial^{2}\sqrt{\rho_{i}}}{\partial
x^{2}}\right)-\beta\frac{\partial a}{\partial x}.
\end{gather}
In the right hand side of \eqref{11} $a(x,t)$ plays the role of
external potential, and the f\/irst term is the derivative of the so
called Bohm potential,
$\frac{1}{2}\frac{1}{\sqrt{\rho_{i}}}\frac{\partial^{2}\sqrt{\rho_{i}}}{\partial
x^{2}}$, and contains all the dif\/fraction ef\/fects (quantum ef\/fects
in quantum problems). By a series of transformations the equation
\eqref{11} is written as~\cite{Fed}
\begin{gather}\label{12}
-\rho_{i}\frac{\partial v_{i}}{\partial t}+v_{i}\frac{\partial
\rho_{i}}{\partial t}+2\left[c_{i}(t)-\int\frac{\partial
v_{i}}{\partial t}~dx\right]~\frac{\partial \rho_{i}}{\partial x}
+\frac{1}{4}\frac{\partial^{3}\rho_{i}}{\partial
x^{3}}-\beta\rho_{i}\frac{\partial}{\partial x}a - 2 \beta a
\frac{\partial \rho_{i}}{\partial x}=0,
\end{gather}
where $c_{i}$ are arbitrary integration constants with respect to
$x$, eventually time dependent. Although~\eqref{12} seems to be
more complicated then the initial equation~\eqref{10}, it can be
solved in two special situations, namely
\begin{itemize}\itemsep=0pt
\item motion with constant velocities $v_{1}=v_{2}=v_{0}$,
 \item motion with stationary prof\/ile current velocity, when all the
quantities $\rho_{i}(x,t)$, $v_{i}(x,t)$, $a(x,t)$ are depending on
$x$ and $t$ through the combination $\xi=x-u_{0}t.$
\end{itemize}
Both cases will be analyzed in the following.

\subsection[Motion with constant velocity ($v_{1}=v_{2}=v_{0}$)]{Motion with constant velocity ($\boldsymbol{v_{1}=v_{2}=v_{0}}$)}

In this case from the continuity equations \eqref{8} one sees that
both $\rho_{1}(x,t)$ and $\rho_{2}(x,t)$ depend on $\xi=x-v_{0}t$.
We assume that also $a(x,t)$ depends only on $\xi $. Then the $a$
equation gives
\begin{gather}\label{13}
a=-\mu(\rho_{1}+\rho_{2}), \qquad \mu=\frac{\gamma}{v_{0}}
\end{gather}
and the equations \eqref{12} write
\begin{gather}\label{14}
\frac{1}{4}\frac{d^{3}\rho_{i}}{d\xi^{3}}-E_{i}\frac{d\rho_{i}}{d\xi}
+2\mu\left(\rho_1 + \rho_2 \right)\frac{d\rho_i}{d\xi} + \mu\rho_i
\frac{d}{d\xi}\left(\rho_1 + \rho_2\right)= 0,
\end{gather}
where by $E_{i}$ we denoted $-(2c_{i}-v_{0}^{2})$,
$\beta\mu\rightarrow \mu$. We shall discuss f\/irstly the situation
$E_{1}=E_{2}, $ the discussion of the more general case $E_{1}\ne
E_{2}$ being postponed for the next subsection. Then the equations
\eqref{14} becomes
\begin{gather}\label{15}
\frac{1}{4}\frac{d^{3}\rho_{i}}{d\xi^{3}}-E\frac{d\rho_{i}}{d\xi}+\mu\rho_{i}\frac{d}{d\xi}(\rho_{1}+\rho_{2})+2\mu
(\rho_{1}+\rho_{2})\frac{d\rho_{i}}{d\xi}=0.
\end{gather}
These are exactly the equations obtained in the case of Manakov's
model~\cite{Man} and extensively discussed by us in \cite{Gre,VGFN}. In the following we shall present several periodic
and traveling wave solutions of~\eqref{15}.

It is convenient to introduce the quantities
$z_{+}=\rho_{1}+\rho_{2}$, and $z_{-}=\rho_{1}-\rho_{2};$ they
satisfy the following equations
\begin{gather}
\frac{1}{4}\frac{d^{3}z_{+}}{d\xi^{3}}-E\frac{dz_{+}}{d\xi}+\frac{3}{2}\mu\frac{dz^{2}_{+}}{d\xi}=0, \nonumber\\
\frac{1}{4}\frac{d^{3}z_{-}}{d\xi^{3}}-E\frac{dz_{-}}{d\xi}+\mu
z_{-}\frac{dz_{+}}{d\xi}+2\mu
z_{+}\frac{dz_{-}}{d\xi}=0.\label{16}
\end{gather}
The second equation is a linear dif\/ferential equation for $z_{-}$
once $z_{+}$ is known. A special solution is
\[
z_{-}=(p_{1}^{2}-p_{2}^{2})z_{+}, \qquad p_{1}^{2}+p_{2}^{2}=1,
\]
which together with the def\/inition of $z_{+}$ gives
\begin{gather*}
\rho_{1}=p_{1}^{2}z_{+}, \qquad \rho_{2}=p_{2}^{2}z_{+}
\end{gather*}
and the problem is reduced in this simple case to f\/ind a solution
of $z_{+}$-equation. This integrated twice gives
\begin{gather}
\label{15-eq} \frac{1}{4}\left(\frac{dz_{+}}{d\xi}\right)^{2}=-\mu
z^{3}_{+}+Ez^{2}_{+}+Az_{+}+B=P_{3}(z_{+}),
\end{gather}
where $P_{3}(z_{+})$ is a third order polynomial in $z_{+}$. The
periodic solution of this equation are easily expressed through
Jacobi elliptic functions.

For constant velocities, as it is easily seen from \eqref{11}, the
densities $\rho_{i}$ have to satisfy the additional conditions
\[
\frac{1}{2}\frac{1}{\sqrt{\rho_{i}}}\frac{\partial^{2}\sqrt{\rho_{i}}}{\partial
x^{2}}+\mu z_{+}(\xi)=\lambda_{i},
\]
which for the previous solutions are satisf\/ied if $\lambda = E/2$
and $B=0$.

Now let us assume that the third order polynomial $P_{3}(z_{+})$
has three distinct roots
\[
P_{3}(z_{+})=-\mu(z_{+}-z_{1})(z_{+}-z_{2})(z_{+}-z_{3}).
\]
The restriction $B=0$ means that one of the roots $z_{2}$ or
$z_{3}$ is  zero. We are interested in positive solutions of~\eqref{16} for which $P_{3}(z_{+})$ is also positive. The periodic
solutions of~\eqref{15-eq} can be expressed through Jacobi
elliptic functions and taking into account the positivity
requirement mentioned before we identify two acceptable situations~\cite{BF}
\begin{gather}
 z_{1}>0, \qquad z_{2}=0,\qquad z_{3}<0,\qquad
 \label{23}
z_{+}=z_{1}\, {\mathrm{cn}}^{2}u,
\\
\nonumber
u=\frac{2\sqrt{\mu}}{g}\xi, \qquad k^{2}=\frac{z_{1}}{z_{1}+\vert
z_{3}\vert }, \qquad g=\frac{2}{\sqrt{z_{1}+\vert z_{3}\vert }}
\end{gather}
and
\begin{gather}
 z_{3}=0, \qquad 0<z_{2}<z_{1}, \qquad
 \label{24}
z_{+}=z_{1}-(z_{1}-z_{2}){\mathrm{sn}}^{2}u,
\\
\nonumber
u=\frac{2\sqrt{\mu}}{g}\xi, \qquad k^{2}=\frac{z_{1}-z_{2}}{z_{1}},
\qquad g=\frac{2}{\sqrt{z_{1}}}.
\end{gather}
Solitary wave solutions are obtained in the limiting case $k=1$
when ${\mathrm{cn}}\, u \rightarrow {\mathrm{sech}}\, u$, ${\mathrm{sn}}\, u \rightarrow {\mathrm{tanh}}\, u$, and both solutions
\eqref{23} and \eqref{24} become a bright soliton
\[
z_{+} \rightarrow z_{1} \frac{1}{\cosh^{2}u}, \qquad
u=\frac{2\sqrt{\mu}}{g}, \qquad g=\frac{2}{\sqrt{z_{1}}}.
\]

It is clear that in this case no energy transfer between the two
components takes place.

The phase $\theta(x,t)$ is easily calculated writing
$\theta_{i}(x,t)=v_{0}x+\gamma_{i}(t)$; then using~\eqref{10} we
get
\[
\theta_{i}=v_{0}x-\left(\frac{1}{2}v_{0}^{2}-\frac{E}{2}\right)t+\delta_{i}.
\]

As far as the f\/ield $a$ is concerned, the solitary wave solution
is
\[
a(u) = -\frac{\mu z_1}{\cosh^2 u}
\]
and, for $\mu z_1\leq u_0$, the f\/ield $\Psi_3$ describes a grey
solution (the inequality gives the dark solution).

\subsection{Motion with stationary-prof\/ile current velocity}

In the case when all the functions depend only on $\xi=x-u_{0}t$
integrating the continuity equa\-tion~\eqref{8} we get
\[
v_{i}(x,t)=u_{0}+\frac{A_{i}}{\rho_{i}},
\]
with $A_{i}$ some integration constants. It is easily seen that
the equations of motion keep the same form as \eqref{14} with
$E_{i}=-(2c_{i}+u^{2}_{0}).$

For $E_1 = E_2$, the results are almost the same as those of the
previous subsection.

However, for $E_1 \neq E_2$, the system \eqref{14} will be solved
by using a direct method. To this end, we look for solutions of
the form
\[
\rho_{i}=A_{i}+B_{i}\,{\mathrm{sn}}\, u, \qquad u=2\lambda\xi,
\]
with $A_{i}$, $B_{i}$, $\lambda$ constants to be determined.
Introducing into \eqref{14} we get \cite{Gre}
\begin{gather}\label{31}
B_{1}+B_{2}=-\frac{4\lambda^{2}k^{2}}{\mu},
\\
-[4\lambda^{2}(1+k^{2})+E_{i}]
B_{i}+\mu(B_{1}+B_{2})A_{i}+2\mu(A_{1}+a_{2})B_{i}=0.\nonumber
\end{gather}
Def\/ining the following new quantities $a_i$, $b_i$, $e_0$ and
$\delta$, i.e.,
\begin{gather*}
B_{i}=-\frac{4\lambda^{2}k^{2}}{\mu}b_i,
\qquad A_{i}=\frac{4\lambda^{2}k^{2}}{\mu} a_{i} ,\\
E_{1}=4\lambda^{2}k^{2}(e_0+\delta) ,
\qquad E_{2}=4\lambda^{2}k^{2}(e_0-\delta) , \qquad \delta >0
\end{gather*}
the f\/irst equation \eqref{31} gives
\[
b_{1}+b_{2}=1,
\]
while from the second, after a little algebra, we get
\begin{gather}
 a_{1}=\frac{1}{3}\left( e_0+\frac{1+k^{2}}{k^{2}}+\delta+4\delta (1-b_{1})\right)b_{1} ,\nonumber \\
 a_{2}=\frac{1}{3}\left(e_0+\frac{1+k^{2}}{k^{2}}-\delta-4\delta (1-b_{2})\right)b_{2} .\label{35}
\end{gather}
As it is expected this result verify the symmetry condition
$1\leftrightarrow 2$ if $\delta\leftrightarrow -\delta$.

Several restrictions result from the positiveness of $\rho_{i}$.
If both $b_{i}$ are positive quantities smaller than unity this
requirement implies
\begin{gather}\label{36}
a_{i}>b_{i}>0.
\end{gather}
Introducing the notation
\[
p=\frac{1}{3}\left(e+\frac{1+k^{2}}{k^{2}}-5\delta\right)
\]
the condition \eqref{36} is satisf\/ied if $p > 1$. In the limiting
case $k^{2}=1$ the solutions are
\begin{gather*}
 \rho_{1}=\frac{4\lambda^{2}}{\mu}\big(a_1-b_1\tanh^{2}u\big),\qquad
 \rho_{2}=\frac{4\lambda^{2}}{\mu}\big(a_2-b_2\tanh^{2}u\big)
\end{gather*}
representing shifted bright solitons.

We can now calculate the f\/ield $a$ from \eqref{13} and using the
limit $a(u)\rightarrow 0$ as $u\rightarrow \infty$, we get $a_1 +
a_2 =1$. Furthermore, making use of \eqref{35}, this condition
becomes
\begin{gather*}
    b_1 - b_2 = \frac{1}{\delta}\left(2 - e_0 -
    \frac{1}{k^2}\right) ,
\end{gather*}
where we have used the condition $b_1 + b_2 =1$.

\section{One bright-two dark solitons interaction}\label{section4}

The previous discussion is easily extended to the situation when
two waves have normal dispersion and one is anomalous,
corresponding to one bright-two dark solitons interaction. Such
vector solitons of mixed bright-dark types are of interest in
quasi-one-dimensional Bose--Einstein condensates (see \cite{PBKK}
and references therein). We are interested in a SW-LW resonance
regime, so several restrictions on the model of three interacting
waves will be imposed. With the notations of section 2 we shall
assume $\left(\frac{\partial \omega_{2}}{\partial
k_{2}}\right)_{0}=\left(\frac{\partial \omega_{3}}{\partial
k_{3}}\right)_{0}$ and after the corresponding translation of
coordinates instead of system \eqref{2} we get
\begin{gather*}
i\frac{\partial A_{1}}{\partial t}+iv\frac{\partial
A_{1}}{\partial
x}+\frac{\alpha_{1}}{2}\frac{\partial^{2}A_{1}}{\partial
x^{2}}+\alpha_{2}\vert A_{1}\vert^{2}A_{1}+\alpha_{3}\vert
A_{2}\vert^{2}A_{1}+\alpha_{4}\vert A_{3}\vert^{2}A_{1}=0,\\
i\frac{\partial A_{2}}{\partial
t}+\frac{\beta_{1}}{2}\frac{\partial^{2} A_{2}}{\partial
x^{2}}+\beta_{2}\vert A_{1}\vert^{2}A_{2}+\beta_{3}\vert
A_{2}\vert^{2}A_{2}+\beta_{4}\vert A_{3}\vert^{2}A_{3}=0,
\\
i\frac{\partial A_{3}}{\partial
t}+\frac{\gamma_{1}}{2}\frac{\partial^{2} A_{3}}{\partial
x^{2}}+\gamma_{2}\vert A_{1}\vert^{2}A_{3}+\gamma_{3}\vert
A_{2}\vert^{2}A_{3}+\gamma_{4}\vert A_{3}\vert^{2}A_{3}=0,
\end{gather*}
where $v=\left(\frac{\partial \omega_{1}}{\partial
k_{1}}\right)_{0}-\left(\frac{\partial \omega_{2}}{\partial
k_{2}}\right)_{0}$ and the rest of the notations are the same as
in Section~\ref{section2}.

Further on we shall consider the channels 2 and 3 with normal
dispersion and the channel 1 with anomalous one and introduce new
f\/ield variables by
\[
A_{1}=\Psi e^{i\delta t},
\qquad A_{2}=(u_{2}+a_{2}(x,t))e^{i(\Gamma_{2}t+\phi_{2}(x,t))}, \qquad A_{3}=(u_{3}+a_{3}(x,t))e^{i(\Gamma_{3}t+\phi_{3}(x,t))}
\]
with $u_{2}$, $u_{3}$,
$a_{2}(x,t)$, $a_{3}(x,t)$, $\phi_{2}(x,t)$, $\phi_{3}(x,t)$ real
quantities, and
\[
\delta=\alpha_{3}u_{2}^{2}+\alpha_{4}u_{3}^{2},
\qquad \Gamma_{2}=\beta_{3}u_{2}^{2}+\beta_{4}u_{3}^{2},
\qquad \Gamma_{3}=\gamma_{3}u_{2}^{2}+\gamma_{4}u_{3}^{2}.
\]
The $A_{1}$-equation transforms into
\begin{gather}\label{40}
i\frac{\partial \Psi}{\partial t}+iv\frac{\partial\Psi}{\partial
x}+\frac{\alpha_{1}}{2}\frac{\partial^{2}\Psi}{\partial
x^{2}}+\alpha_{2}\vert
\Psi\vert^{2}\Psi+2(\alpha_{3}u_{2}a_{2}+\alpha_{4}u_{3}a_{3})\Psi
+\big(\alpha_{3}a_{2}^{2}+\alpha_{4}a_{3}^{2}\big)\Psi=0.
\end{gather}
To get the relevant equations for $a_{2}$, $a_{3}$ we have proceed
like in Section~\ref{section2}, obtaining the following equations for the
amplitudes $a_{2}$, $a_{3}$, i.e.,
\begin{gather}
\frac{\partial^{2}a_{2}}{\partial
t^{2}}+\frac{\beta_{1}}{2}u_{2}\frac{\partial^{2}\vert\Psi\vert^{2}}{\partial
x^{2}}+\beta_{1}u_{2}\left(\beta_{3}u_{2}\frac{\partial^{2}
a_{1}}{\partial
x^{2}}+\beta_{4}u_{3}\frac{\partial^{2}a_{3}}{\partial
x^{2}}\right)+\frac{\beta_{1}^{2}}{4}\frac{\partial^{4}a_{2}}{\partial
x^{4}}+\mbox{(nl.~terms)}  =0,\nonumber\\
 \label{41}
\frac{\partial^{2}a_{3}}{\partial
t^{2}}+\frac{\gamma_{1}}{2}u_{3}\frac{\partial^{2}\vert\Psi\vert^{2}}{\partial
x^{2}}+\gamma_{1}u_{3}\left(\gamma_{3}u_{2}\frac{\partial^{2}
a_{2}}{\partial
x^{2}}+\gamma_{4}u_{3}\frac{\partial^{2}a_{3}}{\partial
x^{2}}\right)+\frac{\gamma_{1}^{2}}{4}\frac{\partial^{4}a_{3}}{\partial
x^{4}}+\mbox{(nl.~terms)}  =0,
\end{gather}
where, as mentioned before the parenthesis $(\cdots)$ of each equation
group all the nonlinear terms which will be irrelevant in a
multiple scales analysis. Furthermore we shall consider only the
special case
$u_{2}=u_{3}=u_{0}$, $\alpha_{2}=\alpha_{3}$, $\beta_{2}=\beta_{3}$, $\gamma_{2}=\gamma_{3}$
and neglect the last nonlinear terms in~\eqref{40} and all the
parenthesis in~\eqref{41}, remaining with the following system of
coupled equations
\begin{gather}
 i\frac{\partial \Psi}{\partial t}+iv\frac{\partial\Psi}{\partial x}+
 \frac{\alpha_{1}}{2}\frac{\partial^{2}\Psi}{\partial x^{2}}+
 \alpha_{2}\vert \Psi\vert^{2}\Psi+2u_{0}\alpha_{3}(a_{2}+
 a_{3})\Psi =0, \nonumber \\
\frac{\partial^{2}a_{2}}{\partial t^{2}}+\frac{\beta_{1}}{2}u_{0}\frac{\partial^{2}
\vert\Psi\vert^{2}}{\partial x^{2}}+\beta_{1}\beta_{3}u_{0}^{2}\frac{\partial^{2}}{\partial x^{2}}
(a_{2}+a_{3})+\frac{\beta_{1}^{2}}{4}\frac{\partial^{4}a_{2}}{\partial x^{2}}=0,\label{42} \\
\frac{\partial^{2}a_{2}}{\partial
t^{2}}+\frac{\gamma_{1}}{2}u_{0}\frac{\partial^{2}\vert\Psi\vert^{2}}{\partial
x^{2}}+\gamma_{1}\gamma_{3}u_{0}^{2}\frac{\partial^{2}}{\partial
x^{2}}(a_{2}+a_{3})+\frac{\gamma_{1}^{2}}{4}\frac{\partial^{4}a_{3}}{\partial
x^{2}}=0.\nonumber
\end{gather}
As in $\Psi$-equation appears the combination $a_{+}=a_{2}+a_{3}$
it is convenient to add the $a_{2}$- and $a_{3}$-equations. In a
multiple scales analysis the fourth order derivatives will
contribute to higher order so we can drop them at the present
stage. Then the system~\eqref{42} reduces to
\begin{gather}
 i\frac{\partial \Psi}{\partial t}+iv\frac{\partial \Psi}{\partial x}+
 \frac{\alpha_{1}}{2}\frac{\partial^{2}\Psi}{\partial x^{2}}+\alpha_{2}
 \vert\Psi\vert^{2}\Psi +2u_{0}\alpha_{3}a_{+}\Psi =0, \nonumber \\
\frac{\partial^{2} a_{+}}{\partial
t^{2}}+\frac{u_{0}}{2}(\beta_{1}+\gamma_{1})\frac{\partial^{2}
\vert\Psi\vert^{2}}{\partial
x^{2}}+u_{0}^{2}(\beta_{1}\beta_{3}+\gamma_{1}\gamma_{3})\frac{\partial^{2}a_{+}}{\partial
x^{2}}=0.\label{43}
\end{gather}
In the resonance condition $(\beta_{1}<0$, $\gamma_{1}<0)$
\[
v=c,
\qquad c^{2}=u_{0}^{2}(\vert\beta_{1}\vert\beta_{3}+\vert\gamma_{1}\vert\gamma_{3})
\]
and by using the multiple scales analysis
\begin{gather*}
x\rightarrow \sqrt{\epsilon} (x-ct), \qquad t\rightarrow \epsilon t,\qquad
a_{+}\rightarrow \epsilon a_{+}, \qquad \Psi \rightarrow
\epsilon^{\frac{3}{4}}\Psi
\end{gather*}
the system \eqref{43} becomes the well known one-component ZYO
system
\begin{gather}
i\frac{\partial \Psi}{\partial t}+\frac{\alpha_{1}}{2}\frac{\partial^{2}\Psi}{\partial x^{2}}+2u_{0}\alpha_{3}a_{+}\Psi =0, \nonumber \\
\frac{\partial a_{+}}{\partial
t}+\frac{u_{0}}{4c}(\vert\beta_{1}\vert +\vert\gamma_{1}\vert
)\frac{\partial \vert\Psi\vert^{2}}{\partial x}=0.\label{45}
\end{gather}
The one-soliton solution of \eqref{45} is given by
\begin{gather*}
a_+ = M {\mathrm{sech}}^{2}\theta, \qquad \Psi =N e^{i\phi}{\mathrm{sech}}\, \theta,\qquad
\theta=\mu(x-x_{0})-\nu t, \qquad \phi=\xi x-\eta t,
\end{gather*}
where
\begin{gather*}
M=\frac{\alpha}{\beta}\mu^{2}, \qquad N^{2}=\frac{\alpha}{\beta\gamma}\mu\nu,\qquad
\nu=\alpha\xi\mu,
\qquad \eta=\frac{\alpha}{2}\big(\xi^{2}-\mu^{2}\big).
\end{gather*}
The integration of the equation for $a_-$ is straightforward, and
it gives
\begin{gather*}
a_- = M_- {\mathrm{sech}}^{2}\theta, \qquad
M_- = \frac{\mu^2}{\nu^2}\left[\frac{u_0}{2}\left(|\beta_1| -
|\gamma_1|\right)N + u_0^2\left(|\beta_1|\beta_3 -
|\gamma_1|\gamma_3\right)M\right] .
\end{gather*}

\section{Conclusions and remarks}\label{section5}

In this paper, an investigation on the interaction between two
bright solitons (anomalous dispersion) and one dark soliton
(normal dispersion) has been carried out. The problem has been
reduced to study a two-component one-dimensional
Zakharov--Yajima--Oikawa system in the case of long wave-short wave
resonance, by using the multiple scales analysis. The system
corresponding to the bright solitons has been discussed further
using a Madelung f\/luid description. In particular, periodic
solutions expressed through Jacobi elliptic functions and
stationary solutions obtained when $k^{2}=1$ have been presented
in two simplifying conditions, namely for constant velocity and
for motion with stationary prof\/ile.

Remarkably, the above two-component Zakharov--Yajima--Oikawa
system is completely integrable and multi-solitons solutions can
be found using dif\/ferent methods, as the bi-linear method of
Hirota \cite{OMO}. However, the Madelung f\/luid description seems
to be useful to f\/ind various solutions of a generalized ZYO
system, containing additional nonlinear terms.

Additionally, the case of two dark and one bright solitons has
been also discussed. Here, the use of the multiple scales analysis
has allowed us to reduced our problem to a simple one-component
Zakharov--Yajima--Oikawa system in the resonance condition. In
particular, the one-soliton solution has been found.

Finally, we would like to point out that the discussion for
arbitrary values of the parameters is still open and
investigations in this direction are in progress. Also numerical solutions of the dynamical model of three interacting waves, discussed in the present paper, could reveal several new aspects of the nonlinearity inf\/luence on the system behavior.

\subsection*{Acknowledgements}

Support through CNCSIS program IDEI-571/2008 is acknowledged. The authors are indebted to an anonymous referee for drawing their attention to the paper \cite{MPK}.

\pdfbookmark[1]{References}{ref}
\LastPageEnding

\end{document}